\documentclass[amsmath,amssymb,aps,showkeys,showpacs,twocolumn,prl]{revtex4-1}

\usepackage[english]{babel}
\usepackage{graphicx}
\usepackage{graphics}
\usepackage{amsmath}
\usepackage{dcolumn}
\usepackage{amssymb}
\usepackage{bm}
\usepackage[latin1]{inputenc}
\usepackage{graphicx,color}
\usepackage[caption=false]{subfig} %perminte colocar duas figuras lado a lado
\newcommand{\mat}[1]{\mbox{\boldmath{$#1$}}} 
\begin{document}
\title{de Haas van Alphen oscillations for neutral atoms in electric fields }
%%%%%%%%%%%%%%%%%%%%%%%%%%%%%%%%%%%%%%%%%%%%%%%%%%%%%%%%%%%%%%%%%%%%%%%%%%%%%%%%%%%%%%
\author{B. Farias}
\email{bruno.farias@ufcg.edu.br}
\affiliation{Centro de Ci\^encias e Tecnologia Agroalimentar, Universidade Federal de Campina Grande, 58840-000, Pombal, PB, Brazil.}
\author{C. Furtado}
\email{furtado@fisica.ufpb.br} 
\affiliation{Departamento de F\'isica, Universidade Federal da Para\'iba, Caixa Postal 5008, 58051-970, Jo\~ao Pessoa, PB, Brazil.} 
%%%%%%%%%%%%%%%%%%%%%%%%%%%%%%%%%%%%%%%%%%%%%%%%%%%%%%%%%%%%%%%%%%%%%%%%%%%%%%%%%%%%%%

%
\begin{abstract}

The de Haas van Alphen (dHvA) effect is well known as an oscillatory variation of the magnetization of conductors as a
function of the inverse magnetic field and the frequency is proportional to the area of the Fermi surface. Here, we show that an analog effect can occur for neutral atoms with a nonvanishing magnetic moment interacting with an electric field. Under an appropriate field-dipole configuration, the neutral atoms subject to a synthetic magnetic field arrange themselves in Landau levels. Using the Landau-Aharonov-Casher (LAC) theory, we obtain the energy eigenfunctions and eigenvalues as well as the degeneracy of the system. In a strong effective magnetic field regime we present the quantum oscillations in the energy and effective magnetization of a two-dimensional (2D) atomic gas. From the dHvA period we determine the area of the Fermi circle of the atomic cloud.

\end{abstract}
\maketitle
\section{INTRODUCTION}
\label{intro}

In the last few decades, the study of the artificial magnetism for neutral atomic systems have grown \cite{1,2,3,4,5,6,7,8,9}. In these systems, neutral particles interacting with a suitable configuration either of electric \cite{10}, magnetic \cite{11} or laser \cite{12} fields behave themselves as charged particles in presence of a synthetic magnetic field. The quest for artificial magnetism is to realize situations where a neutral particle acquires a geometrical phase when it follows a closed contour. In a seminal work \cite{1}, Aharonov and Casher showed that a particle with a magnetic moment moving in an electric field accumulates a quantum phase which is related with a vector potential. This interaction atom-electric field through a nonvanishing magnetic moment (known as Aharonov-Casher (AC) effect) coincides formally in the nonrelativistic limit with that of minimal coupling, where AC vector potential is determined by the electric field and the direction of the magnetic dipole. Based on the AC effect Ericsson and Sj\"oqvist \cite{10} have demonstrated the existence of a certain field-dipole configuration in which an atomic analog of the standard Landau effect \cite{13} occurs. This result has paved the way for the atomic realization of the quantum Hall effect and Shubnikov-de Haas effect as well as de Haas-van Alphen effect using electric fields.

In the solid-state context, dHvA oscillations have been used to study the shape of the Fermi surface of the clean-enough materials  \cite{14,15,16}. Such use is also possible in the context of atomic gases as an alternative technique to the adiabatic band mapping \cite{17}. However, the atomic dHvA effect is still poorly studied. In \cite{18}, Grenier {\it et al.} have explored the possibility of observing like-dHvA oscillations for a non-interacting gas of fermionic atoms, either by putting the gas in rotation or by using artificial gauge fields. In another recent work we propose an experimental scheme for the realization of the dHvA effect in a 2D ultracold atomic cloud which uses the coupling between the internal states of tripod-type atoms and an appropriate spatially varying laser field arrangement \cite{19}.

In this work, we show that the dHvA effect can be induced in a neutral atomic system by the interaction between atoms with magnetic dipole moment and an electric field (AC interaction). From the Ericsson and Sj\"oqvist theory we describe how a symmetric gauge and consequently a uniform magnetic field can be realized in a 2D atomic gas using a suitable field-dipole configuration. This leads to the LAC quantization. We show that the confinement of the neutral particles in an atomic cloud restricts the magnetic field strength. In view of the fact that Rydberg atoms are very sensitive to electric fields, we consider an atomic gas composed by $ ^{87}\mathrm{Rb}  $ ultracold Rydberg atoms and we calculate the LAC degeneracy for this system. In a regime of strong magnetic field and zero temperature we display the quantum oscillations in the energy and effective magnetization of the gas. Finally, as a result of the dHvA oscillations, we determine the area of the Fermi circle of the atomic cloud.

\section{LANDAU-AHARONOV-CASHER QUANTIZATION}

According to the AC theory the Hamiltonian operator that describes the interaction between a neutral particle with nonvanishing magnetic moment $\mat{\mu}$ and an electric field ${\bf E}$, in the non-relativistic limit, is given by \cite{10}
\begin{equation}\label{hamiltonian}
H
=
\frac{1}{2M} \left({\bf p} - \frac{\mu}{c^2} {\bf n} \times {\bf E} \right)^{2} + \frac{\mu \hbar}{2M c^2} \mat{\nabla} \cdot {\bf E},
\end{equation}
where $ {\bf n} $ is the direction of the magnetic dipole moment $\mat{\mu}$, $M$ is the mass of the particle and $c$ is the speed of light. The Hamiltonian (\ref{hamiltonian}) presents an analogy to the minimal coupling, where ${\bf p} - \frac{\mu}{c^2} {\bf n} \times {\bf E}$ is the homologous of the kinematic momentum for a charged particle in the presence of the magnetic field. In this context, the AC vector potential is defined as
\begin{equation}
{\bf A}_{AC} = \frac{1}{c^2} \left({\bf n} \times {\bf E}\right),
\end{equation} 
and the associated magnetic field as
\begin{equation}
{\bf B}_{AC} = \frac{1}{c^2} \mat{\nabla} \times \left({\bf n} \times {\bf E}\right).
\end{equation} 

In order to obtain the LAC quantization we assume the magnetic dipole moment $\mat{\mu}$ aligned parallel with the direction z, i.e., ${\bf n} = \hat{z}$, and we adopt the following cylindrical electric field configuration
\begin{equation}\label{electric field}
{\bf E} = \frac{\rho_{0}}{2 \epsilon_{0}} r \hat{r},
\end{equation} 
where $\epsilon_{0}$ is the electric vacuum permittivity and $\rho_{0}$ is a uniform volume charge density. 

Then the corresponding AC vector potential becomes
\begin{equation}
{\bf A}_{AC} = \frac{\rho_{0}}{2 \epsilon_{0}} r \hat{\phi},
\end{equation}
and the AC magnetic field takes the form of the uniform magnetic field
\begin{equation}\label{0}
{\bf B}_{AC} = \frac{\rho_{0}}{ \epsilon_{0}} \hat{z}.
\end{equation}
Note that the field-dipole configuration presented above obeys the Ericsson and Sjöqvist conditions for the emergence of an AC analog of the Landau effect, namely: (i) condition for vanishing torque on the dipole,  ${\bf n} \times (\langle{\bf p} - \frac{\mu}{c^2} {\bf n} \times {\bf E}\rangle \times {\bf E}) = 0$, where $\langle  \ \cdot \ \rangle$ denotes expectation value; (ii) conditions for electrostatics; $\partial_{t} {\bf E} = 0$ and $ \mat{\nabla} \times {\bf E} = 0$; (iii) ${\bf B}_{AC} $ uniform.

Under the field-dipole configuration presented above we can write the Schr\"odinger equation for the system, in cylindrical coordinates, as
\begin{eqnarray}\label{1}
-\frac{\hbar^{2}}{2 M} \bigg[\frac{1}{r} \frac{\partial}{\partial r} \left(r \frac{\partial}{\partial r} \right) &&+ \frac{1}{r^2} \frac{\partial^2}{\partial \phi^2} \bigg] \Psi + \bigg[- \frac{i\hbar \omega_{AC}}{2}\frac{\partial } {\partial \phi} 
\nonumber\\&& + \frac{M \omega_{AC}^{2}}{8} r^{2} 
+ \frac{\hbar \omega_{AC}}{2}\bigg] \Psi = E \Psi,
\end{eqnarray}
with the cyclotron frequency
\begin{equation}
\omega_{AC} = \frac{\mu \rho_{0}}{M c^{2} \epsilon_{0}} = \frac{\sigma |\mu \rho_{0}|}{M c^{2} \epsilon_{0}} = \sigma |\omega_{AC}|,
\end{equation} 
for which the sign $\sigma = \pm 1$ describes the revolution direction of the corresponding classical motion. 

Once the coefficients in the differential Eq. (\ref{1}) are independent of the azimuth coordinate $\phi$, the angular momentum $ \hat{L}_{z} = i\hbar \frac{\partial}{\partial \phi}$ is a quantum integral of motion and so the wavefunction
 $\Psi$ can be factorized to separate the variables
\begin{equation}\label{2}
\Psi =  e^{i m \phi} R(r). 
\end{equation} 
Here $e^{i m \phi}$ is the eigenfunction of the operator $ \hat{L}_{z}$, with the eigenvalue $\hbar m$ where $m$ is an integer.

Substituting the solution (\ref{2}) into Eq.(\ref{1}) the Schr\"odinger equation assumes the form 
\begin{eqnarray}\label{3}
\frac{\hbar^{2}}{2 M} \left[\frac{d^2}{d r^2} + \frac{1}{r} \frac{d}{dr} - \frac{m^2}{r^{2}} \right] R 
+ \bigg[&& E - \frac{\sigma  \hbar |\omega_{AC}|}{2} \left(m + 1 \right) \nonumber\\&&
- \frac{M \omega_{AC}^{2}}{8} r^{2} \bigg] R = 0.
\end{eqnarray}

Introducing the dimensionless variable $\xi = \frac{M |\omega_{AC}|}{2 \hbar} r^2$ we rewrite Eq. (\ref{3}) in a dimensionless form
\begin{eqnarray}\label{4}
\left[\xi \frac{d^2}{d\xi^2} + \frac{d}{d\xi}\right] R + \left[- \frac{m^2}{4\xi} + \beta - \frac{\xi}{4} \right] R = 0,
\end{eqnarray} 
where $\beta = \frac{E}{\hbar |\omega_{AC}|} - \frac{\sigma}{2} \left(m + 1 \right) $.

The asymptotic analysis of Eq. (\ref{4}) prompts us to write a solution for $R(\xi)$ as
\begin{equation}\label{5}
R(\xi) = e^{-\xi /2} \xi^{|m|/2} W(\xi).
\end{equation} 

It is instructive to note that in case of the standard Landau problem for free electrons in an uniform magnetic field the solution (\ref{5}) is obtained taking $ r \rightarrow \infty$ (which is to say $\xi \rightarrow \infty$). On the other hand, in our system we consider the particles confined in a 2D atomic cloud which implies that the value of $r$ is limited. In this case, we can obtain the analytical solution (\ref{5}), in the limit of \cite{12, 19, 20, 20B}
\begin{equation}\label{}
\frac{M |\omega_{AC}|}{2 \hbar} \gg 1 \ \mathrm{m^{-2}},
\end{equation} 
which leads to the following restriction on the effective magnetic field strength
\begin{equation}\label{cond1}
B_{AC} \gg \frac{2 \hbar c^{2}}{ |\mu| }.
\end{equation}

By substituting solution (\ref{5}) into Eq. (\ref{4}), one arrives in the confluent hypergeometric equation  
\begin{eqnarray}\label{}
\xi \frac{d^2}{d\xi^2} W(\xi) + \bigg[|m| && + 1  -  \xi  \bigg] \frac{d}{d\xi} W(\xi) \nonumber\\ 
&&+ \left[ \beta - \frac{|m| +1}{2} \right] W(\xi) = 0,
\end{eqnarray} 
which is satisfied by the confluent hypergeometric function
\begin{equation}\label{}
W(\xi) =  F \left(-\beta + \frac{|m| + 1}{2}, |m| + 1, \xi \right),
\end{equation} 
with the energy eigenvalues in the form of
\begin{equation}\label{}
E_{n_{\xi}, m}^{(\sigma)} =  \hbar |\omega_{AC}| \left(n_{\xi}  + \frac{|m|}{2} + \frac{\sigma m}{2} + \frac{\sigma}{2} + \frac{1}{2} \right).
\end{equation} 
Here $n_{\xi}$ is a nonzero negative integer. These levels are equivalent to the Landau levels of the
charged system. The radial eigenstates for these LAC states are given by
\begin{eqnarray}\label{}
R_{n_{\xi}, m} (r) =  \frac{1}{a_{AC}^{|m|+1}} && \sqrt{\frac{|m| + n_{\xi}!}{2^{|m|} n_{\xi}! |m|!^{2}} } e^{-\frac{\rho^2}{4 a_{AC}^{2}}} \rho^{|m|} \nonumber\\ && \times F \left(-n_{\xi}, |m| + 1,\frac{\rho^2}{2 a_{AC}^{2}} \right),
\end{eqnarray} 
where $a_{AC} = \sqrt{\frac{\hbar}{M \omega_{AC}} }$.

Finally, introducing a new quantum number 
\begin{equation}\label{}
n =  n_{\xi}  + \frac{|m| + \sigma m}{2},
\end{equation} 
then the energy spectrum acquires the standard form of the LAC spectrum
\begin{equation}\label{}
E_{n}^{(\sigma)} =  \hbar |\omega_{AC}| \left(n + \frac{1}{2}  (1 + \sigma) \right),
\end{equation} 
where $ n = 0,1,2,... $.

The quantities that characterize this LAC system can be obtained by using the AC duality \cite{1,10}
\begin{equation}\label{}
q\Phi \leftrightarrow \frac{\mu  \lambda}{c^{2} \epsilon_{0}},
\end{equation} 
where $\Phi $ is a magnetic flux and $\lambda$ is an uniform linear charge density in the direction of the magnetic dipole. In addition, we have that $\lambda =  \rho_{0} A$, with $A$ being the area of the atomic cloud. 

In this way, the spacing between the energy levels for a fixed $\sigma$ is $ \Delta E_{AC}= \hbar \frac{|\mu \rho_{0}|}{M c^{2} \epsilon_{0}} $, the effective magnetic length is $ l = \sqrt{\frac{\hbar c^{2} \epsilon_{0}}{\mu \epsilon_{0}}} $
and the degeneracy is
\begin{equation}\label{degeneracy}
D_{AC} =  \rho B_{AC}
\end{equation} 
where $ \rho = \frac{\mu A}{c^{2} h} $ and $ B_{AC} = \frac{\rho_{0}}{ \epsilon_{0}}$. Note that the degeneracy linearly depends on the magnetic field $B_{AC}$ and it is limited by the fact that we have a finite atomic trap.

From an experimental point of view, ultracold Rydberg atomic clouds are good candidates for the realization of the LAC quantization. This is because the extreme properties of the highly excited atoms compared to atoms in ground state, such as very high dipole polarizabilities, magnetic moments and atom-atom strengths become the Rydberg atoms very sensitive to electric fields \cite{21,22,23}. For instance, if an $ ^{87}\mathrm{Rb}  $ atom is excited to a state $ n = 51$ ($n$ is the principal quantum number), the atom can have a magnetic moment of $ \mu = 50 \ \mu_{B}$, which is a factor of $50$ higher than the magnetic moment of an atom in a $ | 5S_{1/2}, F=2, m_{F} = 2 \rangle$ ground state \cite{23}. This would lead to a stronger AC interaction for the Rydberg atom in an electric field, as compared to an atom in the ground state. Furthermore, the regime of ultracold temperatures allows reaching a significant LAC quantization without the requirement of too extreme electric fields. 

In what follows we consider a 2D ultracold atomic cloud with an area of $ A \sim 150 \ \ \mathrm{\mu m^{2}} $ and contains $ N \sim 10^{4} $ $ ^{87}\mathrm{Rb}  $ atoms in the $ n = 51$ excited state \cite{24,25,26}. Under these conditions, the expression (\ref{cond1}) becomes
\begin{equation}\label{}
B_{AC} \gg 40.93 \ \ \mathrm{T_{eff}},
\end{equation}
where $\mathrm{T_{eff}} = \ \mathrm{\frac{N}{C \cdot m}}$ is an effective unit.
 
In addition, the degeneracy is written as
\begin{equation}\label{}
D_{AC} = 1.17 \times 10^{-15}  B_{AC}.
\end{equation}
As a result, the lowest landau level regime of the system, i.e., $ D_{AC} = 10^{4} $, is achieved when $ B_{AC} = 8.55\times 10^{18} \ \ \mathrm{T_{eff}} $.

%%%%%%%%%%%%%%%%%%%%%%%%%%%%%%%%%%%%%%%%%%%%%%%%%%%%%%%%%%%%%%%%%%%%%%%%%%%%%%%%%%%%%%%%%%
\section{Quantum oscillations for neutral particles subject to an electric field}

From now on, we investigate the phenomenon of dHvA oscillations for the neutral atomic system presented in the previous section. We consider the system at zero-temperature limit and containing a fixed number of $ N $ atoms. We do not take into account the temperature smearing of the quantum oscillations. We assume that the lowest $p$ LAC levels (where $p$ is a positive integer) are completely filled with $ p D $ atoms each one and the highest $ (p+1)\mathrm{th} $ level is partly occupied with $ N - p D $ atoms. In this case, the Fermi level lies in the $ (p+1)\mathrm{th} $ level.

\begin{figure}[!ht]
	\begin{center}
		\subfloat[]{\label{Number_atom}\includegraphics[scale= 0.8]{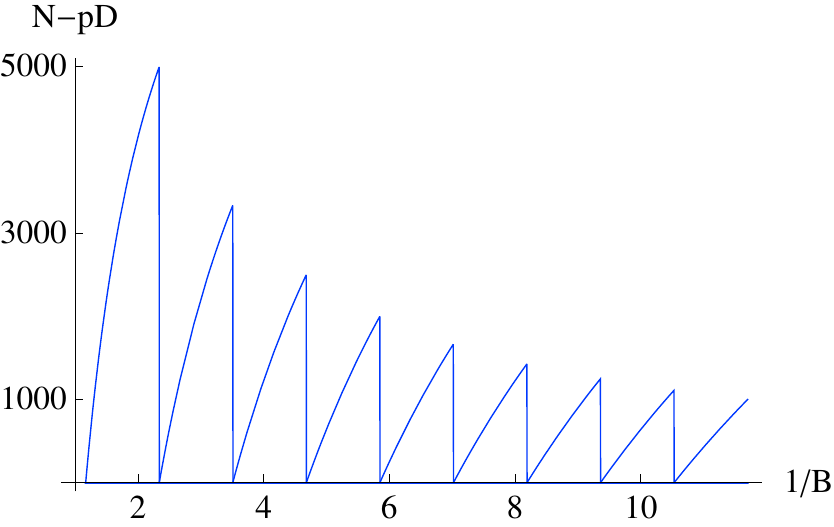}}
		\hspace{0.5cm}\subfloat[]{\label{Number_atomII}\includegraphics[scale= 0.8]{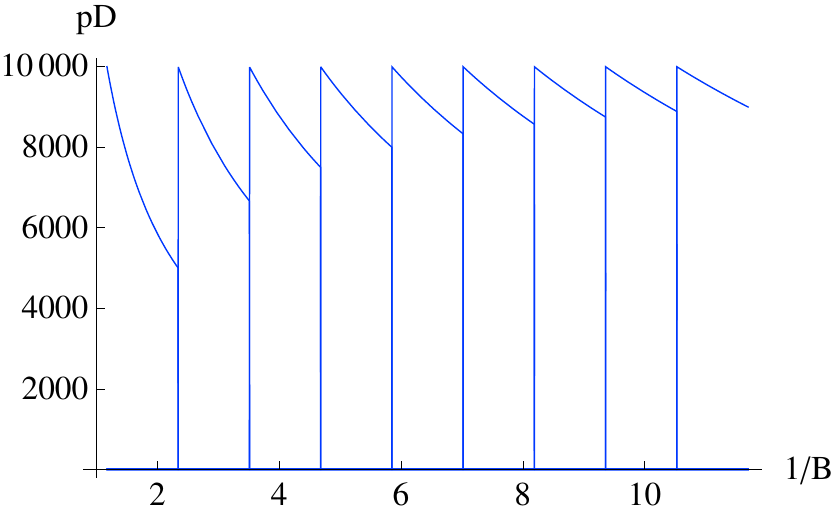}}
		\end{center}
\caption{(a) (Color online) Number of $ N - p D $ atoms in the LAC level partially filled as a function of the reciprocal magnetic field (in units of $ 10^{-19}  \ \mathrm{T_{eff}^{-1}} $).  
     (b) Number of $ s D $ atoms in energy levels which are completely occupied as a function of the inverse magnetic field.
}
\end{figure}

From Eq. (\ref{degeneracy}) we can see that by decreasing the magnetic field $B_{AC}$,  it decreases the degeneracy $D$ of the system. As a consequence, fewer atoms can be accommodated on each LAC level and the atomic population of the highest $ (p+1)\mathrm{th} $  energy level will range from completely full to entirely empty. Fig.(\ref{Number_atom}) illustrates the transfer of atoms to the highest partly occupied LAC level of a 2D ultracold cloud with $ N = 10^{4} $ $ ^{87}\mathrm{Rb}  $ atoms when the magnetic field is slept in the range $ 8.55\times 10^{18} \ \ \mathrm{T_{eff}} \leq {B_{AC}} \leq   8.55\times 10^{17} \ \ \mathrm{T_{eff}} $ or equivalently $ 1.17 \times 10^{-19} \ \mathrm{T_{eff}^{-1}} \leq \frac{1}{B_{AC}} \leq   1.17 \times 10^{-18} \ \mathrm{T_{eff}^{-1}} $. At $ \frac{1}{B_{AC}} = 1.17 \times 10^{-19} \ \mathrm{T_{eff}^{-1}} $ only the lowest level $ p = 0 $ is populated with $ 10^{4} $  atoms and the upper level $ (p+1)\mathrm{th} = 1 $ is empty. As the reciprocal magnetic field is increased the level $ p = 1 $ starts to accommodate atoms until it is fully occupied with $ 5 \times 10^{3} $ atoms. Then a new upper level $ (p+1) \mathrm{th} = 2 $ becomes populated and so on. Note that singularities appear at strength of the magnetic field where a new LAC level becomes occupied. The distance between such singularities are regularly spaced and defines the dHvA period $ \Delta(\frac{1}{B_{AC}}) = \frac{\rho}{N} = 1.17 \times 10^{-19}  \ \mathrm{T_{eff}^{-1}} $. Fig.(\ref{Number_atomII}) displays how the population of the lowest $p$ completely filled levels varies with $ \frac{1}{B_{AC}} $. It is this jump of atoms to a higher energy level that causes the oscillations in the energy and in the effective magnetization of the atomic gas as a function of the inverse AC magnetic field. 

Summing the energy of the atoms in fully occupied levels with the energy of the atoms in the partly filled level we have the total energy of the system that is given by
\begin{equation}\label{energy3}
 \varepsilon =\sum_{n=0}^{p-1}(n+\frac{1}{2})\hbar|\omega_{AC}|D + (N - p D)\hbar|\omega_{AC}|(p+\frac{1}{2})
\end{equation}
which can be rewritten in a most appropriate form as
\begin{equation}\label{energy3}
\varepsilon =-\mu^{\mathrm{eff}}_{B}\rho \left( B_{AC}^2 p(p+1) - \frac{N}{\rho} B_{AC}(2p+1)\right),
\end{equation}
for $\frac{\rho s}{N} < \frac{1}{B_{AC}} < \frac{\rho (s + 1)}{N} $. Here $ \mu^{\mathrm{eff}}_{B} = \frac{\hbar \mu}{2M c^{2}} $ is an effective Bohr magneton. 

\begin{figure}[h!]
        \centering
        \includegraphics[scale=0.9]{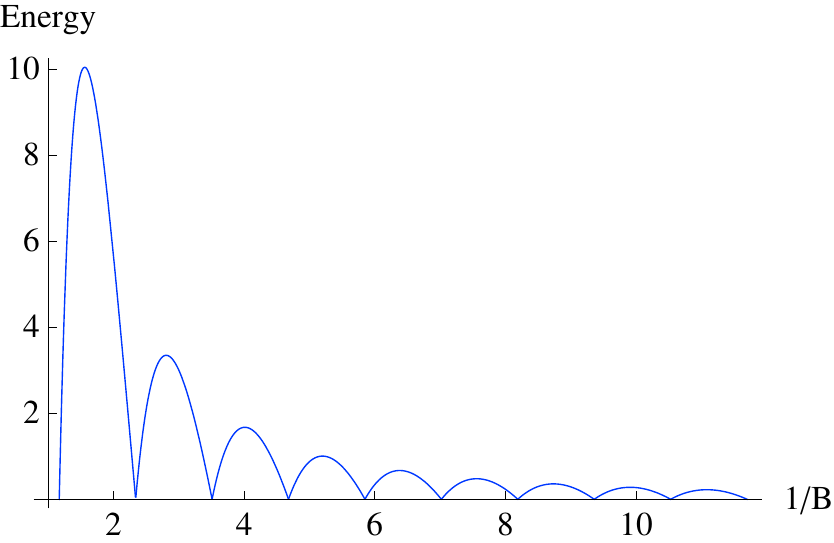}       
        \caption{
		Variation of the energy (in units of $ 10^{-27}\ \ \mathrm{J} $) of the partially occupied LAC level with respect to the inverse effective magnetic field (in units of $ 10^{-19}\ \ \mathrm{T_{eff}^{-1}} $). For calculations, we take $ M_{\mathrm{Rb}} = 1.443\times10^{-25}\ \mathrm{kg}$ and $ \mu = 4.64 \times 10^{-22}\ \ \mathrm{J/T}  $.
}
\label{energy}
\end{figure}

However, for simplicity, we consider just the total energy of the atoms in the partly occupied LAC level which is expressed as
\begin{equation}\label{energy4}
\varepsilon' = - \mu^{\mathrm{eff}}_{B}\rho\left[ p B_{AC}  - \frac{N}{\rho}\right] \left[(p+1) B_{AC} - \frac{N}{\rho} \right]. 
\end{equation}
As Fig. (\ref{energy}) shows, the energy $\varepsilon'$ oscillates with a quadratic dependence on each dHvA period.  Note that if we set either $ \frac{1}{B_{AC}^{p}} = \frac{p\rho}{N} $ or $ \frac{1}{B_{AC}^{p+1}} = \frac{(p+1)\rho}{N} $ there is no LAC level partly populated and the Eq. (\ref{energy4}) is zero. On the other hand, at $ \frac{1}{B_{AC}} = \frac{\rho}{N}\frac{p(p+1)}{p+\frac{1}{2}} $ the energy $\varepsilon'$ has a local maximum. The amplitude of the oscillations falls because a new higher level passes to be occupied each time with fewer particles as $ \frac{1}{B_{AC}} $ varies.

The effective magnetization $\mathcal{M} $ is obtained taking $ - \frac{\partial \varepsilon'}{\partial B_{AC} } $, then we have 
\begin{eqnarray}\label{magnetization}
\mathcal{M} =  \mu^{\mathrm{eff}}_{B}\rho\left[ 2B_{AC} p(p+1) - \frac{N}{\rho} (2p+1) \right], 
\end{eqnarray}
where $ pD < N \leq (p+1)D $.
In Fig. (\ref{magnetization}) we display the dHvA oscillations of the AC magnetization as a function of the inverse magnetic field calculated from Eq. (\ref{magnetization}). As expected, at $T = 0$ these oscillations have a saw-tooth shape with a constant amplitude. $\mathcal{M}$ is linear in $ \frac{1}{B_{AC}} $ except in points for which a new LAC level starts to be filled. In these points the Fermi level makes an abrupt jump between two LAC levels and consequently $\mathcal{M}$ experiences jumps of $ 2 N \mu^{\mathrm{eff}}_{B} \approx 3.76 \times 10^{-44} \ \mathrm{\frac{J \cdot s}{T}} $ at the end of each dHvA period. The effective magnetization is $ - N \mu^{\mathrm{eff}}_{B} $ (Landau diamagnetism) when just the first $ p $ LAC levels are occupied, and it jumps to $ + N \mu^{\mathrm{eff}}_{B} $ as the $ (p+1)$th level starts to be populated, returning smoothly to $ - N \mu^{\mathrm{eff}}_{B} $ again when this new level is completely populated.

It is important to observe that since the AC magnetic field in Eq. (\ref{0}) is artificial, the effective magnetization $\mathcal{M}$ is not directly observable. However, as discussed in \cite{18} there are indications that the dHvA oscillations should be observed in the angular momentum $\langle L_{z} \rangle $ of the atomic gas, which is directly analogous to the magnetization in the solid-state context.

\begin{figure}[h!]
        \centering
        \includegraphics[scale=0.9]{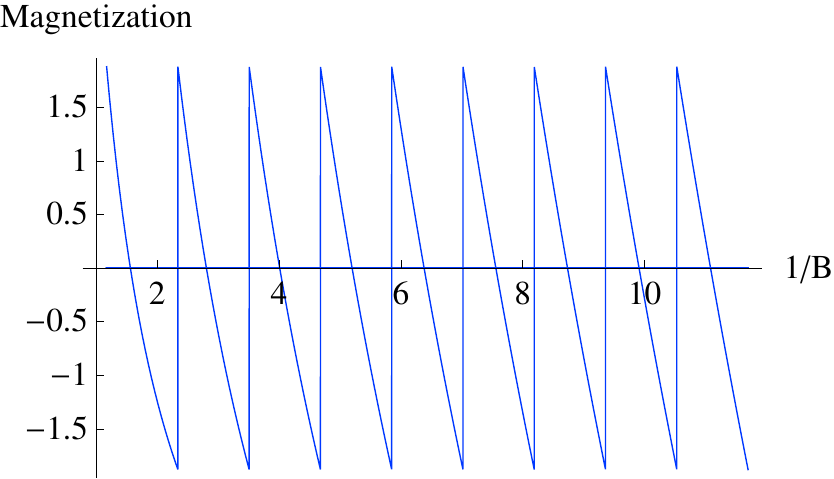}       
        \caption{
		Effective Magnetization (in units of $ 10^{-44}\ \mathrm{\frac{J \cdot  s}{T}} $) as a function of the inverse magnetic field (in units of $ 10^{-19}\ \ \mathrm{T^{-1}} $).
}
\label{magnetization}
\end{figure} 

As a natural consequence of the dHvA oscillations, we can determinate the area $S$ of the extremal cross-section of the Fermi surface of the atomic cloud through the Onsager-like relation
\begin{eqnarray}
S = \frac{2\pi N}{\hbar \rho} \nonumber
\end{eqnarray}
Using the physical parameters of the system we estimate the area of the Fermi circle in $ S \approx 5 \times 10^{53} \ \mathrm{m^{-2}} $.

%%%%%%%%%%%%%%%%%%%%%%%%%%%%%%%%%%%%%%%%%%%%%%%%%%%%%%%%%%%%%%%%%%%%%%%%%%%%%%%%%%%%%%%%%%
\section{Conclusion}

An atomic analog of the de Haas van Alphen effect based on the Aharonov-Casher interaction has been proposed. In this context, neutral atoms may interact with an electric field via a nonvanishing magnetic moment. Using the Ericsson and Sj\"oqvist theory we have shown how a uniform magnetic field can be induced by an electric field in a 2D ultracold atomic gas. We have shown that the magnetic field strength is limited by the fact that the gas is finite. Applying the Schrödinger equation approach the LAC energy levels and eigenfunctions are obtained. Due to the high sensitivity of the Rydberg atoms to the electric fields we have considered an atomic cloud composed by $ ^{87}\mathrm{Rb}  $ ultracold Rydberg atoms and we have calculated the LAC degeneracy for this system. In the limit of high magnetic field and zero temperature we have presented the oscillatory variation of the energy and effective magnetization of the gas as a function of the magnetic field strength. As a consequence of the dHvA oscillations, we have estimated the area of the Fermi surface of the atomic gas.

%%%%%%%%%%%%%%%%%%%%%%%%%%%%%%%%%%%%%%%%%%%%%%%%%%%%%%%%%%%%%%%%%%%%%%%%%%%%%%%%%%%%%%%%%%
\section{Acknowledgments}

We would like to thank CNPq and CAPES   for financial support.


\begin{thebibliography}{99}



\bibitem{1} Y. Aharonov, and A. Casher, Phys. Rev. Lett. {\bf 53}, 319 (1984).

\bibitem{2} J. Anandan, Phys. Rev. Lett. {\bf 48}, 1660  1982.

\bibitem{3} X.-G. He, B.H.J. McKellar, Phys. Rev. A {\bf 47}, 3424 (1993).

\bibitem{4} M. Wilkens, Phys. Rev. Lett. {\bf 72}, 5 (1994).

\bibitem{5} H. Wei, R. Han, X. Wei, Phys. Rev. Lett. 75 (1995) 2071.

\bibitem{6} M. V. Berry, Proc. R. Soc. A {\bf 392}, 45 (1984).

\bibitem{7} A. L. Fetter, Rev. Mod. Phys. {\bf 81}, 647 (2009).

\bibitem{8} J. Ruseckas, G. Juzeli$ \bar{\mathrm{u}} $nas, P. \"Ohberg, and M. Fleischhauer, Phys. Rev. Lett. {\bf 95}, 010404 (2005).

\bibitem{9} Y. J. Lin {\it et al.}, Nature (London) {\bf 462}, 628 (2009).

\bibitem{10} M. Ericsson, E. Sj\"oqvistt, Phys. Rev. A {\bf 65}, 013607 (2001).

\bibitem{11} L.R. Ribeiro, C. Furtado, and J.R. Nascimento, Phys. Lett. A {\bf 348}, 135 (2006).

\bibitem{12} B. Farias, J. Lemos de Melo, and C. Furtado, Eur. Phys. J. D {\bf 68}, 77 (2014).
 
\bibitem{13} L.D. Landau, Z. Phys. {\bf 64}, 629  (1930).

\bibitem{14}  W. de Haas, and P. van Alphen, Comm. Phys. Lab. Leiden {\bf 212a} (1930). 

\bibitem{15} A. Abrikosov, Introduction to the theory of normal metals, Solid State Physics: Supplement (Academic Press,1972).

\bibitem{16} N. Ashcroft, and N. Mermin, Solid State Physics (Saunders College, Philadelphia, 1976).

\bibitem{17} M. K\"ohl, H. Moritz, T. St\"oferle, K. G\"onter, and T. Esslinger, Phys. Rev. Lett. {\bf 94}, 080403 (2005).

\bibitem{18} Ch. Grenier, C. Kollath, and A. Georges, Phys. Rev. A {\bf 87}, 033603 (2013).

\bibitem{19} B. Farias, and C. Furtado, Physica B {\bf 481}, 19 (2016). 

\bibitem{20} P. \"Ohberg, G. Juzeli$ \bar{\mathrm{u}} $nas, J. Ruseckas, and M. Fleischhauer, Phys. Rev. A {\bf 72}, 053632 (2005).

\bibitem{20B} G. Juzeli$ \bar{\mathrm{u}} $nas, and P. \"Ohberg, Phys. Rev. Lett. {\bf 93}, 033602 (2004).
 
\bibitem{21} Jongseok Lim, Han-gyeol Lee, and Jaewook Ahn, Journal of the Korean Physical Society, {\bf 63}, No. 4, 867 (2013).

\bibitem{22} T. Pohl, H. R. Sadeghpour, and P. Schmelcher, Physics Reports {\bf 484} 181 (2009).

\bibitem{23}  R. R. Mhaskar. Toward an Atom Laser: Cold Atoms in a Long, High gradient Magnetic Guide. PhD thesis, University of Michigan, (2008).

\bibitem{24} Another important point to be considered in an experimental arrangement is that the electric field, given by Eq. (\ref{electric field}), requires a uniform charge density inside of the atomic cloud. Although experimentally challenging, such an arrangement could be composed of a 2D toroidal trap and a perpendicular line of charge passing through the centre of the toroid \cite{25}.

\bibitem{25} A. A. Wood, B.H.J. McKellar, and A. M. Martin, arXiv:1604.04996v1.

\bibitem{26} O. Morizot, Y. Colombe, V. Lorent, and H. Perrin and B. M. Garraway,  Phys. Rev. A {\bf 74}, 023617 (2006).
 

\end{thebibliography}
\end{document}